\documentclass[useAMS,usenatbib]{mn2e}

\usepackage{graphicx}

\title[Long-term polarimetric observations of OH127.8+0.0]{Long-term polarimetric observations of OH127.8+0.0}

\author[P. Wolak, M. Szymczak and E. G\'erard]
{P. Wolak$^1$\thanks{E-mail: wolak@astro.uni.torun.pl}, 
M. Szymczak$^{1}$
and E. G\'erard$^{2}$
\\
$^{1}$Toru\'n Centre for Astronomy, Nicolaus Copernicus University, Gagarina 11, 87-100 Toru\'n, Poland \\
$^{2}$GEPI, UMR 8111, Observatoire de Paris, 5 place J. Janssen, 92195 Meudon Cedex, France \\}

\begin{document}

\date{Accepted 2013 January 9. Received 2013 January 9; in original form 2012 October 25}

\pagerange{\pageref{firstpage}--\pageref{lastpage}} \pubyear{2013}

\maketitle

\label{firstpage}
\begin{abstract}
OH 1612- and 1667-MHz masers from the well known object OH127.8+0.0 were monitored in full-polarization mode
over a period of 6.5\,years and mapped with MERLIN at one epoch.
The OH variability pattern of the star is typical of extremely long-period AGB stars. The distance 
determined from the 1612-MHz light curve and a new measurement of the angular radius is 3.87$\pm$0.28\,kpc.
At both frequencies, the flux of polarized emission tightly follows the total flux variations while,
the degrees of circular and linear polarization are constant within measurement accuracy.. There is
net polarization at both lines. The magnetic field strength estimated from a likely Zeeman pair is 
$-$0.6\,mG at the distance of 5400\,au from the star.At the near and far sides of the envelope, the polarization vectors are well 
aligned implying a regular structure of the magnetic field. 
The polarization characteristics of the OH maser emission suggest a radial magnetic field which is 
frozen in the stellar wind.  
\end{abstract}

\begin{keywords}
masers -- polarization -- circumstellar matter -- stars: AGB and post-AGB
\end{keywords}

\section{Introduction}
OH/IR stars have a spectral energy distribution which peaks between 10 and 25$\mu$m,
an absorption feature at 9.7$\mu$m and a double-peaked OH maser profile at 1612\,MHz.
Almost all of them are long period variables with periods of 500--3000 days and
bolometric amplitudes of 1--2 magnitudes \citep{habing96}.
As low and intermediate mass stars with enhanced mass loss up to a few 
$10^{-5}$M$_{\sun}$\,yr$^{-1}$, which leads to a very optically thick and dusty circumstellar 
envelope, they could be considered as successors of Miras (\citealt{vanderveen88}; 
\citealt{vassi93}; \citealt{habing96}). Untargeted surveys of OH 1612-MHz
line suggested that OH/IR stars have progenitor masses from 1 to 8M$_{\sun}$ \citep{baud81}.
OH/IR objects with periods above 1000\,days are thought to be probably medium mass ($\ge$4.5M$_{\sun}$) 
evolved objects (\citealt{likkel89}; \citealt*{lepine95}; \citealt{habing96}).
The study of mid-infrared spectral features in a large sample of circumstellar OH masers
provides indirect evidence for mass segregation between Miras and OH/IR objects \citep{chen01}.
One of the ways to verify this hypothesis and to constrain the evolutionary status of OH/IR
stars is to assess their parameters using reliable distances.

OH127.8+0.0 (= OH127 = IRAS01304+6211 = V669\,Cas) was discovered in a small-scale untargeted 
survey at 1612\,MHz \citep{kerr74}. High angular resolution studies of this 
brightest OH/IR star in the anti-centre region revealed a thin OH 1612-MHz maser shell of 
$\sim$1\farcs5 radius expanding with a velocity of about 12\,km\,s$^{-1}$ (\citealt{booth81}; 
\citealt*{norris82}; \citealt*{bowers83}).
These observational characteristics are fully consistent with the predictions
of the standard expanding shell model (\citealt{goldreich76}; 
\citealt*{elitzur76}), although a large portion of the envelope shows departure from 
spherical symmetry. In this model, a thin spherical shell and double-peaked OH 1612-MHz profiles
are the natural consequence of radial amplification of OH maser emission in a uniform outflow at 
a constant velocity \citep{reid77}.
The star has received much attention over the years as an archetypal OH/IR object but its distance 
is poorly determined spreading from 2.9 to 6.2\,kpc (\citealt*{vanlange90}; 
\citealt{bowers90}). 
In this paper we estimate its distance comparing the linear diameter of the shell, 
determined using the phase lag technique \citep{herman85},
with the angular diameter measured on interferometric maps.

In early observations, the OH maser emission from OH/IR stars appeared largely unpolarized
\citep{cohen89}. However, our recent study has revealed polarized features in about 
60\% (33/57) of the OH/IR stars \citep*{wolak12}. Elliptically polarized emission
with a degree of polarization up to 30\% is usually detected. This indicates the presence of
magnetic fields in the outer envelopes. High angular resolution observations have
 revealed a magnetic 
field of a few mG, whose structure was imaged in several late-type stars (\citealt{chapman86};
\citealt*{szymczak98}, 2001; \citealt{bains03}; \citealt*{amiri10}).
OH127 has been mapped several times with the VLA (\citealt{bowers83}; 
\citealt{bowers90}) and MERLIN (\citealt{booth81}; 
\citealt{norris82}) but no full polarization data were obtained.  
We present polarimetric maps of the target and analyze full polarization time series in order
to explore the behaviour of the magnetic field and the degree of saturation of maser amplification.

\section{Observations}
\subsection{Nan\c{c}ay}
Data were obtained at 189 epochs during the period 2002 May 12 $-$ 2008 December 28 
(Table \ref{maintable}) with the Nan\c{c}ay radio telescope (NRT). 
A cooled 1.1$-$1.8\,GHz receiver and a 8192-channel autocorrelation
spectrometer were used for simultaneous observations of the two OH transitions at 1612.231 and 1667.359\,MHz. 
The receiver system temperature was about 35\,K and the effective velocity resolution was 0.142 and
0.137\,km\,s$^{-1}$ at 1612 and 1667\,MHz, respectively. The half-power beam width was 3\farcm5(RA)
by 19\arcmin(Dec) and the conversion factor of the antenna temperature to the flux density for a point
source was $\sim$1.4\,K\,Jy$^{-1}$. 

The two orthogonal linear polarizations and two opposite circular polarizations for each transitionwere measured in the 8-bank mode of the 
 spectrometer. The Stokes parameters $I$, $Q$ and $V$ were directly
provided by the system, while the Stokes $U$ was extracted by a horn rotation of 45\degr.
The four Stokes parameters were used to derive: the linearly polarized flux density, $p =\sqrt{Q^2 + U^2}$,
degree of circular polarization, the $m_{\rm C} = V/I$, degree of linear polarization, $m_{\rm L} = p/I$ 
and the polarization position angle, $\chi = 0.5$tan$^{-1}(U/Q)$.
The absolute and relative accuracies of gain measured with a noise diode were $\sim$5\% and $\sim$1\%,
respectively.
The error in the polarized intensity caused by the polarization leakage between the orthogonal feeds was about 2\%.
 The baseline subtraction was done by frequency switching: this mode introduces an error of less than 0.6\% 
in the polarization parameters, as compared to the position switching mode.
W12 (an unpolarized source in absorption) and W3OH (a strongly polarized maser source) were regularly observed
throughout the whole project in order to measure the instrumental polarization and consistency of the amplitude 
calibration.
The absolute flux density scale was accurate to within 7$-$8\%.
A detailed description of the methods of observations, full polarization calibrations, analysis of
errors is given in \citet{szymczak04}. A total integration time for the two 
horn positions of $\sim$12\,min resulted in a $3\sigma_{\rm rms}$ sensitivity level in the Stokes $I$ of 
$\sim$0.4\,Jy. 
All the velocities given in this paper are relative to the local standard of rest.

 Since the upgrade of the NRT in 2001 \citep*{vandriel96} interferences from local and/or satellite transmitters were reduced
dramatically. Throughout the whole observing period, we noticed them at a few epochs only and the total loss of data was
less than one per cent.

\subsection{MERLIN}
OH127 was observed in the 1612- and 1667-MHz OH transitions on 2001 December 21 using the six telescopes
of the MERLIN array in full-polarization mode. The narrow-band (0.25\,MHz) observations of the target 
switched at intervals of 3 minutes between the two maser lines and were interleaved with 1.5-min scans
on 0141$+$579, the phase reference source, in wide-band mode (14\,MHz) at the appropriate frequencies
to obtain the optimum signal-to-noise ratio. The source 3C84 was observed both in narrow-band and wide-band 
modes in order to determine the phase offset correction that arises due to observations made 
in different frequency setups and to derive corrections for instrumental gain variations across the bandpass.
The flux density scale and the absolute polarization position angles were determined by observations 
of 3C286.  

The data reduction was carried out following standard procedures for MERLIN OH polarimetry
\citep{diamond03} using the d-programs at Jodrell Bank and the Astronomical Image 
Processing System (AIPS). The $I$, $Q$, $U$ and $V$ data cubes were obtained with a beam of 
0\farcs17$\times$0\farcs15. The spectral channel width was 0.5\,kHz, 
which corresponds to 0.10 and 0.09\,km\,s$^{-1}$ at 1612 and 1667\,MHz respectively. The 1$\sigma_{\rm rms}$ 
noise level in emission-free Stokes $I$ single channel map was typically 17\,mJy\,beam$^{-1}$.
The error on the absolute flux scale is 5 per cent. The errors on the polarization angle were usually lower 
than $\pm$2\fdg0 for the strongly polarized components. The absolute maser position error was $\pm$20\,mas.

\section{Results}

\subsection{Total flux density curves}
\begin{figure}   
\resizebox{\hsize}{!}{\includegraphics[angle=-90]{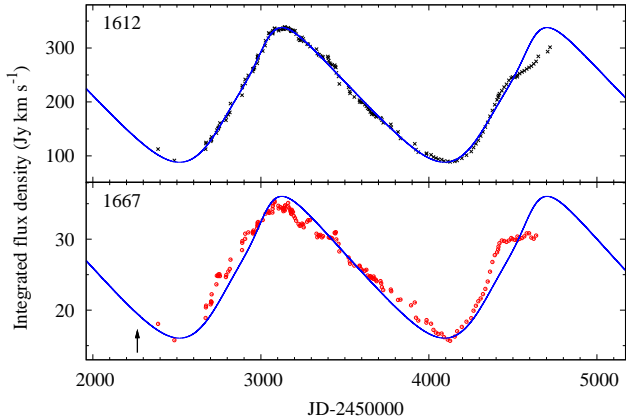}}
\caption{Integrated flux density of the OH 1612- (upper panel) and 1667-MHz (lower panel) spectra of OH127 as
         a function of time. The uncertainties of the measurements are comparable with the symbol sizes. 
         The blue line shows the model light curve. The arrow marks the date of the MERLIN polarimetric observation.  
\label{lightcurves}}
\end{figure}

The variations of the total integrated flux densities at 1612 and 1667\,MHz are shown in Figure \ref{lightcurves}.
In order to represent the light curve shape a sum of two periodic functions of the type discussed by \citet*{david96}
was fitted to the data. The time of maximum for the red-shifted integrated flux density, $T_0$, the period, $P$, 
the amplitude measured in the logarithmic scale, $\Delta m$, and the asymmetry defined as the fraction of the period 
between minimum and maximum are given in Table \ref{maintable}. 
There are no significant differences in those parameters between the 1612 and 1667-MHz lines with the exception of 
$\Delta m$, which differs by a factor of 1.7. The ratios of the integrated flux at the two lines range from 5.4 at
minimum light to 9.9 at maximum. This implies a partial suppression of the 1667-MHz emission as the brightness
increases due to a competitive gain effect \citep{gray95}.  

The total flux density at the two frequencies shows smooth and regular variations with the exception of a bump on 
the rising branch of the light curve near MJD = JD$-$2450000 = 4450. A similar event occurred near JD = 2446800 
as might be deduced from the 1612-MHz curve shown by \citet{vanlange90} (their Fig. 2).
The asymmetry deduced from our data is higher than that reported previously (\citealt{herman85}; 
\citealt{vanlange90}). 
The period inferred from our data is shorter by 434 days than reported in \citet{herman85}. 
It is also shorter by 38 days than the value given in \citet{vanlange90} but is well within 
their uncertainties. An ephemeris of the time of maximum based on Herman \& Habing and van Langevelde et al.'s data
is inconsistent with that derived from our observations covering about one and half cycles. The large errors 
in their estimates cannot account for that inconsistency and we suggest significant changes of the light curve of 
OH127 from one cycle to the next. 
We conclude that the OH maser curves of OH127 strongly resemble the general characteristics of the optical light curve 
of long-period ($<$600 days) and extremely long-period ($\sim$1850 days) Miras (\citealt{lebzelter11}; 
\citealt{groenewegen09}).  

\begin{table}
\caption{Measured and derived parameters for OH127.8+0.0. \label{maintable}}
\begin{tabular}{l c}
\hline
 &  \\
\multicolumn{2}{c}{\it Nan\c{c}ay data} \\
 &  \\
Span of monitoring data (MJD)    & 2355 $-$ 4735 \\
Number of observations           &  189  \\
Expansion velocity, $V_{\rm e}(1612)$ (km\,s$^{-1}$) & $11.13 \pm 0.14$  \\
Expansion velocity, $V_{\rm e}(1667)$ (km\,s$^{-1}$) & $11.31 \pm 0.14$  \\
Stellar velocity, $V_*$ (km\,s$^{-1}$)   & $-54.78\pm0.14$  \\
Time of maximum, $T_0(1612)$ (MJD)      & 3164.5$\pm$2.5 \\
Time of maximum, $T_0(1667)$ (MJD)      & 3086.1$\pm$20.2 \\
Period, $P(1612)$ (days)                &  1599.8$\pm$4.1 \\
Period, $P(1667)$ (days)                &  1596.1$\pm$4.8 \\
Amplitude, $\Delta m(1612)$ (mag)     & 1.416$\pm$0.011  \\
Amplitude, $\Delta m(1667)$ (mag)     & 0.829$\pm$0.013  \\
Asymmetry of the light curve (1612)     & 0.406$\pm$0.006  \\
Asymmetry of the light curve (1667)     & 0.343$\pm$0.013  \\
Phase lag, $\Delta\Phi(1612)$ (days)  & 61.81$\pm$1.10 \\
Shell radius (au)         &   $5358.3 \pm 95.7$   \\
 &  \\
\multicolumn{2}{c}{\it MERLIN data} \\
 &  \\
RA offset of the shell centre$^a$ (mas) & $0.0$  \\
Dec offset of the shell centre$^a$ (mas) & $-191.0$ \\
Outer shell radius, $r_{\rm o}(1612)$ (mas)   & $1380\pm40$  \\
Inner shell radius, $r_{\rm i}(1612)$ (mas)   & $1150\pm40$  \\
Expansion velocity, $V_{\rm e}(1612)$ (km\,s $^{-1}$) &  $11.7\pm0.2$  \\
Outer shell radius, $r_{\rm o}(1667)$ (mas)   & $1340\pm70$  \\
Inner shell radius, $r_{\rm i}(1667)$ (mas)   & $1120\pm70$  \\
Expansion velocity, $V_{\rm e}(1667)$ (km\,s $^{-1}$) &  $11.4\pm0.2$  \\
Stellar velocity, $V_*$ (km\,s $^{-1}$)         &  $-54.95\pm0.20$ \\
Distance (kpc) &  $3.87\pm0.28$   \\

\hline

\multicolumn{2}{l}{$^a$Relative to the phase centre of }\\
\multicolumn{2}{l}{RA(J2000) = 01$^{\rm h}$33$^{\rm m}$51\fs232, 
Dec(J2000) = 62$^0$26\arcmin53\farcs236}\\

\end{tabular}
\end{table}

\subsection{Phase lag}
\begin{figure}   
\resizebox{\hsize}{!}{\includegraphics[angle=-90]{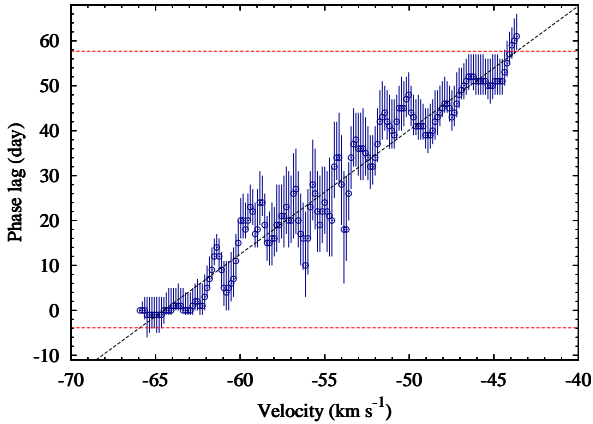}}
\caption{Phase lag between the light curves of spectral channels in the 1612-MHz spectrum
         of OH127 plotted against the velocity. The dashed black line is the best least squares 
         fit to the whole data set. The red lines cross the best fit line at the velocities
         of the strongest blue- and red-shifted peaks. \label{phaselag}}
\end{figure}

The time shifts between the 1612-MHz flux curves of individual channels, i.e. phase lags are determined 
using the following method. The light curves of each channel were normalized and smoothed with a three-point
running mean and linearly interpolated to obtain uniformly one-day spaced data. The channel with the strongest 
emission at $-$65.9\,km\,s$^{-1}$ was chosen as a reference, i.e. as the zero point of the phase lag.
All the other channels were scaled in amplitude and shifted in time to minimize the least squares difference 
between their light curves and that of the reference channel. 
The phase lags of all channels are plotted in Figure \ref{phaselag}. The linear relation between the phase
lag and velocity is fully consistent with the model of a uniformly expanding shell. The phase lag between the
strongest blue- and red-shifted peaks, $\Delta\Phi$, is 61.81$\pm$1.10\,days (Table \ref{maintable}). 
Although the light curve is not strictly the same for each stellar cycle, the interpolation and smoothing applied 
to the data do not affect the phase lag because our observations were highly sampled with a typical time 
resolution of two weeks ($\sim$0.009$P$). 

Our value of $\Delta\Phi$ differs significantly from that of 75.6$\pm$3.6 days
and 41.4$\pm$8.1 days reported by \citet{herman85} and by \citet{vanlange90},
respectively. Such a drastic difference between these previous estimates of the phase lag is likely to be due to 
their different methods as, the two teams used a similar data sets. 
Our method is similar to that of van Langevelde et al., 
 but the data have a much higher temporal sampling.
We confirm an asymmetry 
of the OH shell that may cause an error in the phase lag measurements of up to 15\% \citep{bowers90}.

\subsection{Polarization spectra and light curve}
\begin{figure*}   
\resizebox{\hsize}{!}{\includegraphics[angle=-90]{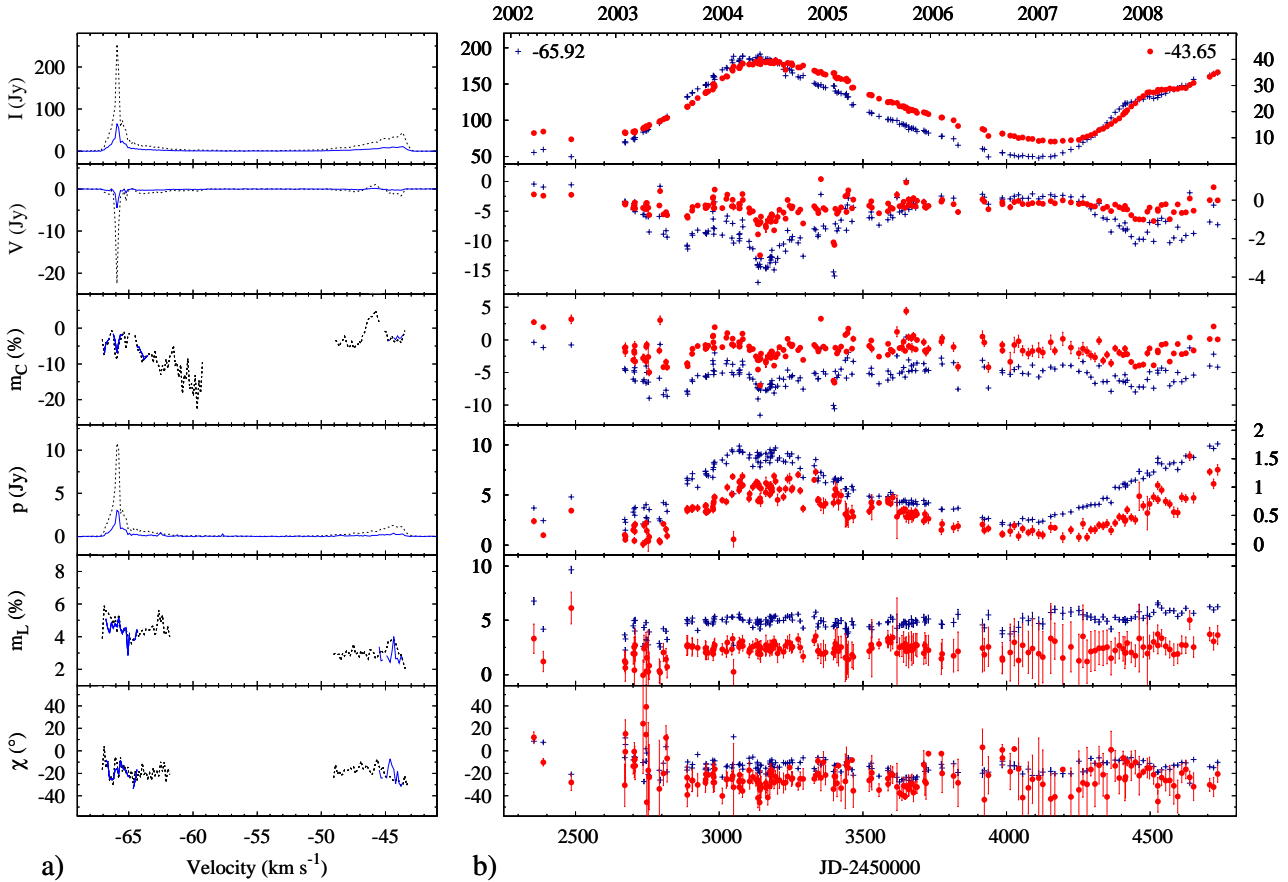}}
\caption{Polarization spectra of OH127 at 1612\,MHz. ({\bf a}) The dashed and solid lines are the average spectra
  for MJD ranges 3072$-$3203 (27 observations) and 4013$-$4168 (9 observations) near the maximum and minimum of the light curve, 
  respectively. The Stokes $I$, $V$, degree of circular polarization, $m_{\rm C}$, linearly polarized flux density, $p$,
  degree of linear polarization, $m_{\rm L}$, and polarization position angle, $\chi$, are shown from top 
  to bottom. ({\bf b}) Time-series of the polarimetric parameters for the extreme blue- ($-$65.92\,km\,s$^{-1}$)
  and red-shifted ($-$43.65\,km\,s$^{-1}$) emission peaks. For the Stokes $I$, $V$ and $p$ the left- and right-hand ordinates 
  give the flux density scales for the extreme blue- and red-shifted peaks, respectively. 
 Error bars are shown only for the red-shifted emission peak for sake {\bf of} clarity but those for 
the blue-shifted emission peak are similar. For the  Stokes $I$ values the error bars are generally 
smaller than or of comparable size to the symbols.
\label{spectr1612}}
\end{figure*}

\begin{figure*}   
\resizebox{\hsize}{!}{\includegraphics[angle=-90]{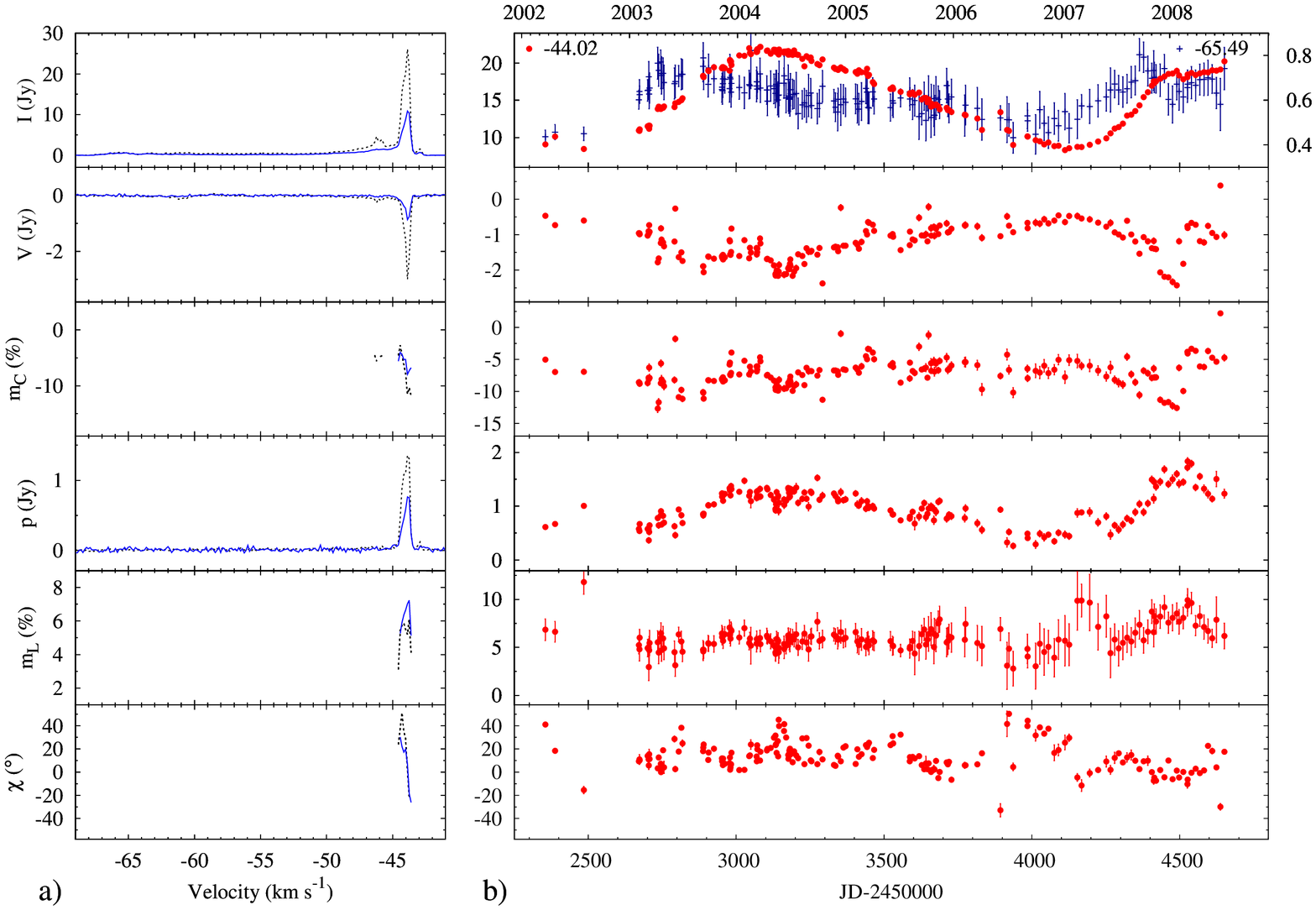}}
\caption{Same as in Fig.\ref{spectr1612} but for 1667-MHz line and with two exceptions: the average spectra are for
  MJD ranges 3081$-$3210 (27 observations) and 4042$-$4224 (9 observations) near the maximum and minimum of the light curve, 
  respectively and that the time-series of all polarimetric parameters are shown for the extreme red-shifted 
  ($-$44.02\,km\,s$^{-1}$) emission peak, while the Stokes $I$ flux density only is shown for the extreme blue-shifted 
  ($-$65.49\,km\,s$^{-1}$) emission peak.
\label{spectr1667}}
\end{figure*}

Figures \ref{spectr1612} and \ref{spectr1667} show the average polarization spectra near the maximum
and minimum and the time series of polarization parameters for the strongest features at 1612 and 1667\,MHz, 
respectively. 

The expansion velocity defined as the half-maximum width in the OH profile peaks at 1612-MHz, $V_{\rm e}$(1612)
does not differ from that at 1667-MHz, $V_{\rm e}$(1667) (Table \ref{maintable}). The stellar velocity determined
as the mean velocity of the outer 1612-MHz peaks, $V_{\rm *}$ (Table \ref{maintable}) and the OH expansion 
velocities are quite consistent with previous estimates \citep{bowers90}. 
The outflow velocity of 13.0\,km\,s$^{-1}$ determined from the observations of CO transitions 
\citep{debeck10} suggests that the OH masers form in regions which still did not 
reach the terminal velocity.

The 1612-MHz profile shows strong left hand circularly (LHC) polarized
 emission over almost the whole velocity
range. The degree of circular polarization is usually lower than 20\%. It generally decreases with an increase of
the total flux density presumably due to depolarization effect. The degree of linear polarization is generally lower 
than 6\% and shows a slight difference between the blue- and red-shifted parts of the spectrum. 
The polarization position angle is very similar in all spectral channels with a mean value of $-$14\fdg7$\pm$1\fdg5.

 The 1667-MHz profile also shows strong LHC emission,
 but a relatively narrow polarized feature is seen only 
in the red-shifted part of the spectrum and $m_{\rm C}$ is lower than 12\%. A linearly polarized feature is also
detected in the red-shifted edge of the emission. The maximum value of $m_{\rm L}$ is 6--7\%, while the $\chi$ angle 
changes rapidly across the profile of 1.1\,km\,s$^{-1}$ width, the mean value being 10\fdg2$\pm$7\fdg6.
We conclude that the quantitative polarization properties of the 1612- and 1667-MHz maser profiles of OH127 are 
typical for OH/IR objects in the sample recently studied \citep{wolak12}.

The timeseries of all polarization parameters of the brightest 1612- and 1667-MHz features at the blue- and red-shifted parts 
of the spectrum (Figs. \ref{spectr1612} and \ref{spectr1667}) exhibit regular variations of the Stokes $V$ and $p$ fluxes tightly 
correlated with the total flux variations. The mean and median values of the degrees of circular and linear polarization 
and the polarization angle of those features are summarized in Table \ref{fracpol}.
The degrees of circular and linear polarization were constant within the uncertainty of instrumental polarization of 
about 2--3\%. The polarization angle also does not show any systematic variations higher than 8--13\degr. 
Outliers in the plots are of instrumental origin, probably due to unbalanced gains of both types of polarization that 
we could not correct using the calibration data available.
We found that the difference between the average polarization position angles for the four brightest blue- and red-shifted
channels of the 1612-MHz profile was 1\fdg0$\pm$13\fdg5. This suggests that the 1612-MHz maser regions are permeated 
with a regular magnetic field that is stable over a $\sim$6.5 year period.
The linearly polarized emission at 1667\,MHz was detected only in the red-shifted part of the profile. 
In the velocity range of $-$43.9 to $-$43.6\,km\,s$^{-1}$ the mean difference of $\chi$ between 1667- and 1612-MHz spectra is only
6\fdg5$\pm$15\fdg3 but it increases to 46\fdg9$\pm$18\fdg4 for the range $-$44.6 to $-$44.0\,km\,s$^{-1}$. 
This indicates nearly the same projected direction of magnetic fields in the compact regions of the far sides of 
the 1612-MHz and 1667-MHz shells, whereas in the annular regions, these transitions likely probe different clouds.   

\begin{table}
 \caption{Mean and median values of the polarization parameters for the brightest OH features derived from the multi-epoch data.\label{fracpol}}
\begin{tabular}{c c r r r r}
\hline
Line  &   Velocity      &   Parameter   &  Mean &   SD$^a$ &  Median \\
(MHz) & (km\,s$^{-1}$)   &               & \multicolumn{3}{c}{} \\
\hline
1612 & $-$65.92 &  $m_{\mathrm C}$(\%)    & $-$5.34  & 1.81  & $-$5.32  \\
     &          &  $m_{\mathrm L}$(\%)    &  5.13  & 1.90  &  5.00  \\
     &          &  $\chi$(\degr)       & $-$13.53  & 6.97  & $-$25.02 \\
     & $-$43.65 &  $m_{\mathrm C}$(\%)    & $-$1.46  & 1.68  & $-$1.41  \\
     &          &  $m_{\mathrm L}$(\%)    &  2.53  & 1.12  &  2.45  \\
     &          &  $\chi$(\degr)       & $-$23.32 & 13.14 & $-$25.03\\    
 1667& $-$44.02 &  $m_{\mathrm C}$(\%)    & $-$7.20  & 2.18  & $-$7.07 \\
     &          &  $m_{\mathrm L}$(\%)    & 6.05    & 1.56   & 5.81 \\
     &          &  $\chi$(\degr)       & 12.95    & 12.86  & 11.82 \\
\hline

\multicolumn{2}{l}{$^a$ Standard deviation}
\end{tabular}
\end{table}

\subsection{Shell parameters and distance}
\begin{figure}   
\resizebox{1.0495\hsize}{!}{\includegraphics[angle=0]{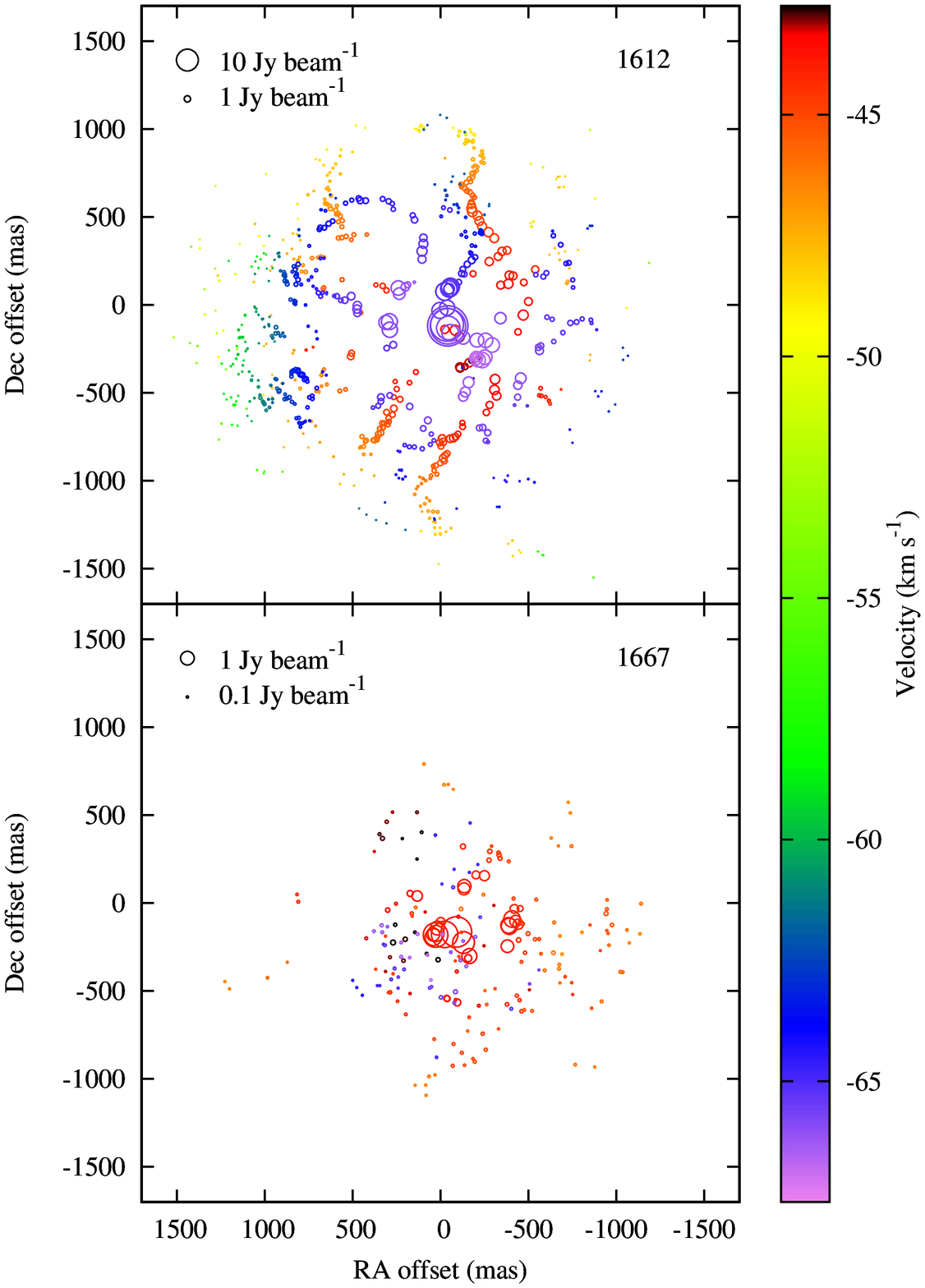}}
\caption{Stokes $I$ maps of the 1612- and 1667-MHz maser spots of OH127. The symbol sizes are proportional to the logarithm 
  of the brightness. The velocity scale is coded in colour.  
\label{StokesI_spots}}
\end{figure}

The maps of OH emission at 1612 and 1667\,MHz are well known (\citealt{booth81}; 
\citealt{norris82}; \citealt{bowers83}; \citealt{diamond85}; \citealt{bowers90}) and we added the Stokes $I$ spot maps 
(Fig. \ref{StokesI_spots}) only for comparison purposes.
The new observation generally confirms the paucity of 1667-MHz emission in the eastern side of the well pronounced
1612-MHz shell. The extent of 1667-MHz emission in the western side is the same as that of 1612-MHz shell, which
indicates that the two masers operate at similar distances from the central star. 

\begin{figure}   
\resizebox{1.0\hsize}{!}{\includegraphics[angle=-90]{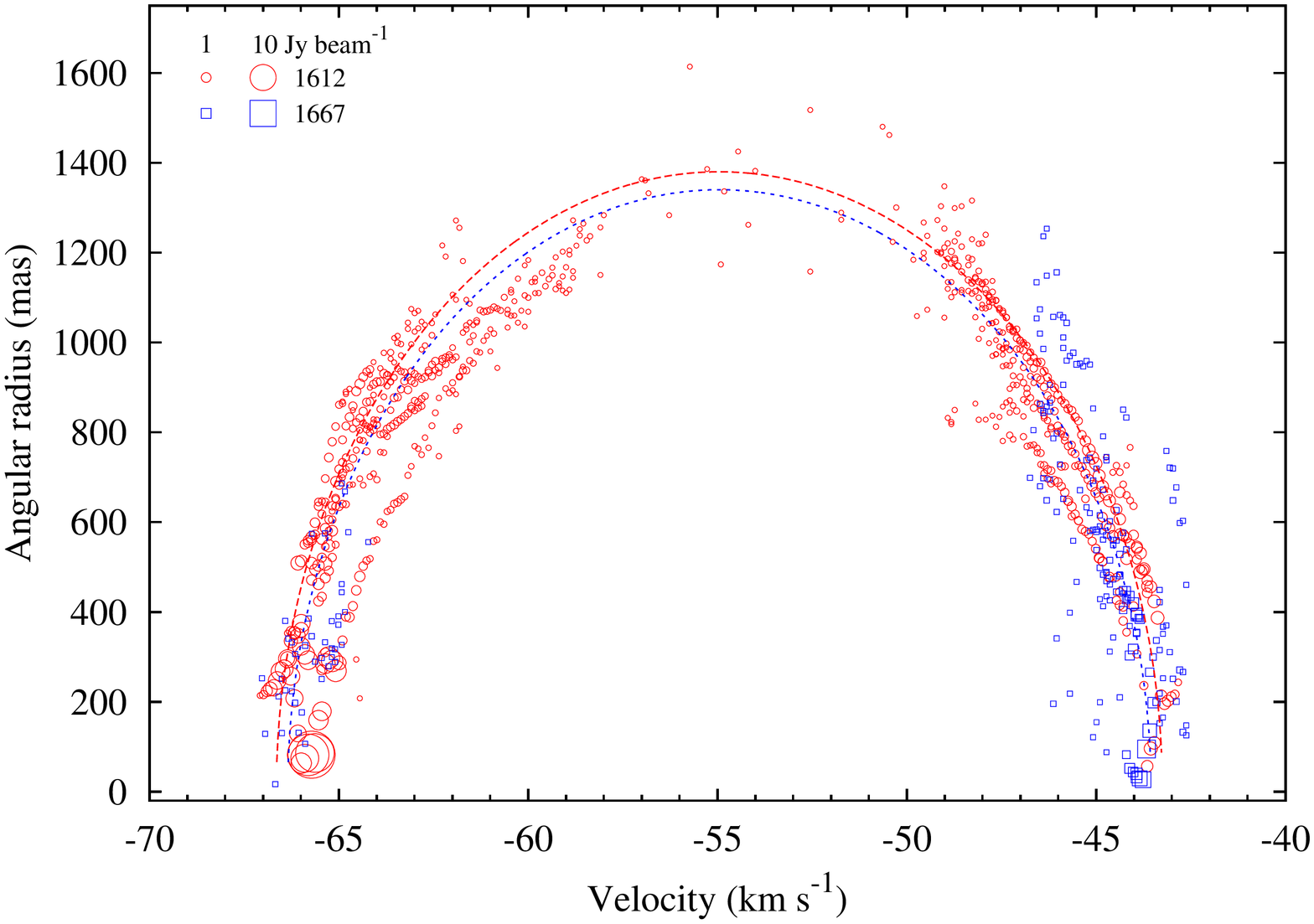}}
\caption{Angular distance of the maser emission spots from the estimated stellar position 
of OH127 versus their radial velocity. The symbol sizes are proportional to the logarithm 
of the brightness. The least-squares fits of the thin shell model to the data are used to estimate the shell parameters
as listed in Table \ref{maintable}. The red-dashed and blue-dotted curves show the outer radii fitted to the maser
distribution at 1612 and 1667\,MHz, respectively.   
\label{diagram}}
\end{figure}

The locations and velocities of the Stokes $I$ maser components at both frequencies obtained from the MERLIN observations
are used to refine the shell parameters: stellar position, expansion and stellar velocities, outer, $r_{\rm o}$ 
and inner, $r_{\rm i}$ shell radius. The standard thin-shell model \citep{reid77} was fitted 
to the Stokes $I$ spots (Fig. \ref{diagram}) using a weighted least-squares algorithm. Parameters for the best fitting model 
are listed in Table \ref{maintable}. These are consistent with those reported by \citet{bowers90}. 

Assuming a spherical thin shell model with isotropic emission, 
the measurement of the 1612-MHz emission extent to the $3\sigma$ level integrated over the velocity range 
from $-$56.0 to $-$54.2\,km\,s$^{-1}$ reveals an OH shell radius of 1343$\pm$30\,mas, which agrees
with the estimate of the standard model (Table \ref{maintable}). 
 The assumption of a spherical thin shell with isotropic emission allows us to consider this radius as the
angular size of the shell. For non-isotropic emission, the angular size would be overestimated 
by $\le$10\%. An increase of shell thickness and maser saturation leads to higher biases \citep{vanlange93}.
Combining the shell radius with the phase lag 
provided by the NRT observations, we obtain a distance of 3.87$\pm$0.28\,kpc for OH127.

\subsection{Spatial structure of the polarized emission}
The polarization data obtained with MERLIN are analyzed using the standard methods 
(e.g \citealt{szymczak98}; \citealt{bains03}). Tables \ref{circular} and \ref{linear} 
list the parameters of the polarized components at both frequencies. 
Only the components with $V$ and $p$ brightness higher than $\sim$100\,mJy\,beam$^{-1}$ are listed.
These make up about 10 and 7\% of all Stokes $I$ components at 1612 and 1667\,MHz respectively.

\begin{figure}   
\includegraphics[angle=0,width=0.5\textwidth]{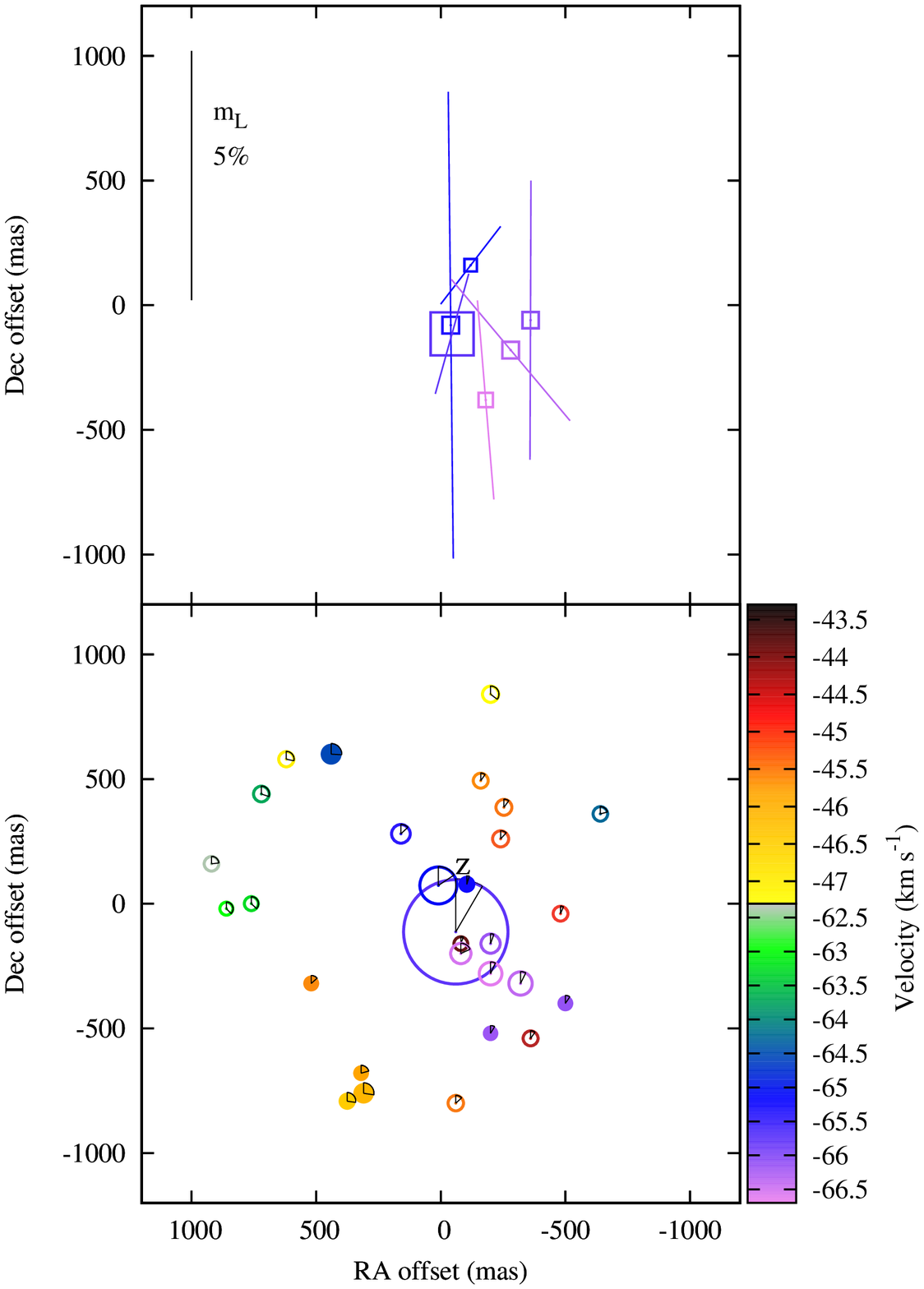}
\caption{Distribution of the 1612-MHz maser polarized components of OH127. Velocities are coded by colours as shown 
         in the bottom right wedge. {\it Upper panel}: linearly polarized components. The size of squares is proportional
         to the logarithm of the $p$ flux density and the length of polarization vectors is proportional to the degree
         of linear polarization, $m_{\mathrm L}$. {\it Lower panel}: circularly polarized components. The size of the circles 
         is proportional to the logarithm of the Stokes $V$ flux density. Empty and solid circles indicate negative and 
         positive values of the Stokes $V$, respectively.The size of sectors is proportional to the degree of circular
         polarization.\label{map1612}}
\end{figure}

The spatial distribution of the 1612-MHz polarized components is shown in Fig. \ref{map1612}. The origin of the maps
is the position of the central star as derived from the standard model (Sect. 3.4). 
All of the 28 detected components have velocities larger than 0.65$-$0.70$V_{\rm e}$ 
with respect to the stellar velocity and their typical Stokes $V$ flux is
0.1$-$0.2\,Jy\,beam$^{-1}$. There is a strong excess of $-V$ components (20/28). The blue-shifted components are much 
closer to the  projected
 position of the central star than the red-shifted components. 
The degree of circular polarization ranges from 5 to 38\% which is a factor of 2
higher than that observed in the NRT spectra implying depolarization in the NRT beam.
Difference in the spectral resolution by a factor of 1.5 between both instruments also plays a role.
There are only a few components with elliptical polarization.  The linearly polarized emission, generally weaker than 
0.5\,Jy\,beam$^{-1}$ is detected from the blue-shifted part of the spectrum. The components with $m_{\rm L} \le$9\%
appear in a cluster of size $\sim$400\,mas located $\sim$200\,mas west of the stellar position. 
The polarization angles range from $-$37 to 40\degr with a mean flux-weighted value of $-$4\fdg6$\pm$12\fdg4 that is
consistent with the NRT results (Sect. 3.3).  

\begin{figure}   
\includegraphics[angle=0,width=0.5\textwidth]{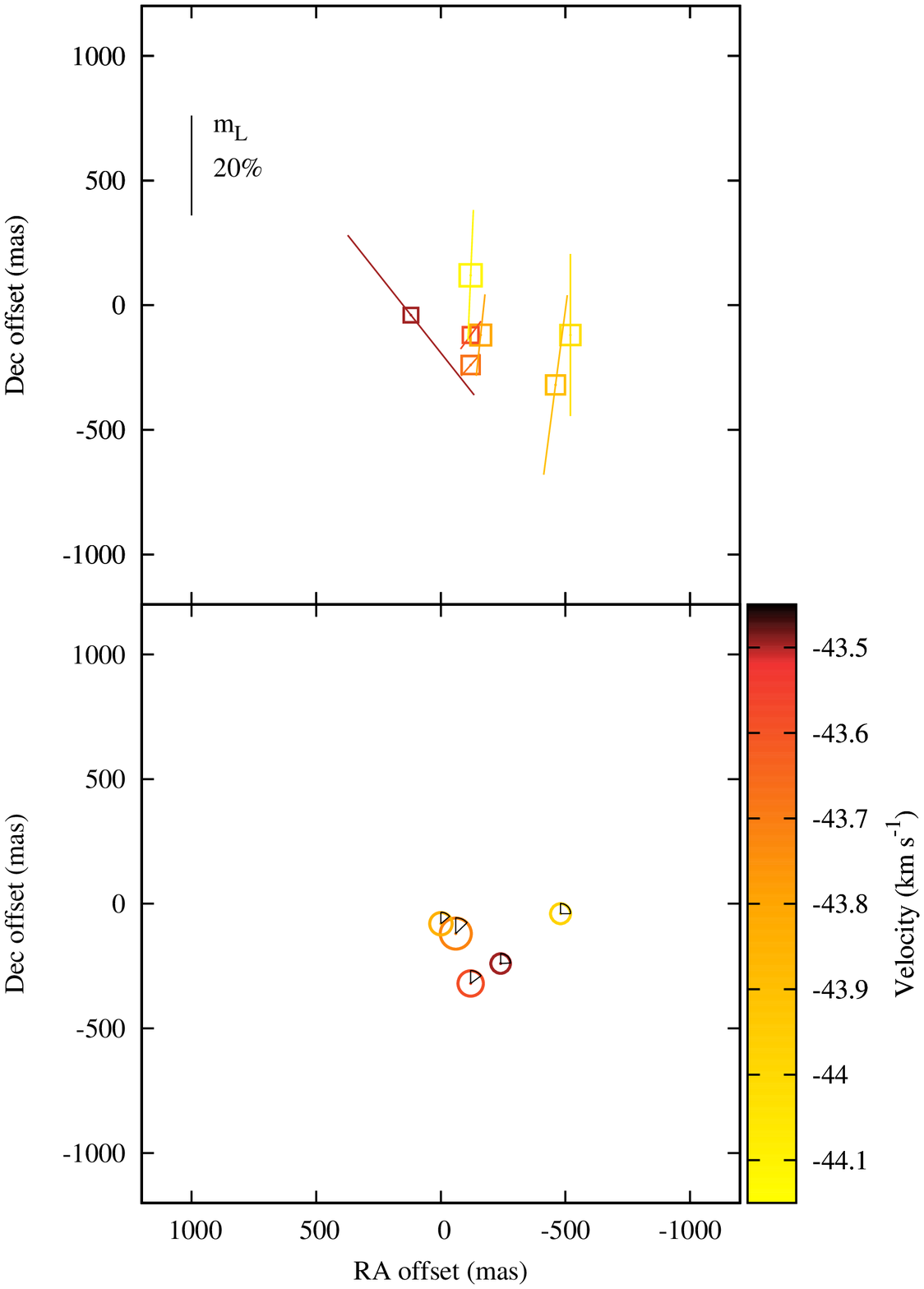}
\caption{Same as Fig. \ref{map1612} but for the 1667-MHz emission.\label{map1667}}
\end{figure}

The polarized 1667-MHz components are seen at the red-shifted velocities. The circularly polarized components of 
flux lower than 0.3\,Jy\,beam$^{-1}$ and $|m{\rm_C}|\le$25\% are clustered near the stellar position (Fig. \ref{map1667}).
All of them appear as LHC components which is consistent with the NRT results. The linearly polarized components show a similar
location as the circularly polarized emission. The mean value of $m_{\rm L}$=23.6$\pm$5.3\% is a factor 5 higher than that
for the 1612-MHz emission. The mean flux-weighted polarization angle is $-$9\fdg5$\pm$11\fdg5. This agrees well with that 
at 1612\,MHz and implies the same projected direction of magnetic fields in the regions of both OH masers. 

While the degrees of circular polarization at 1612 and 1667\,MHz are similar, the degrees of linear polarization 
at 1667\,MHz are much higher than at 1612\,MHz. This can imply that the linear polarization at 1612\,MHz
is suppressed or depolarized in the regions of strong amplification of this line \citep{elitzur96}.

\subsubsection{Magnetic field strength}
\begin{figure}   
\includegraphics[angle=-90, width=0.51\textwidth]{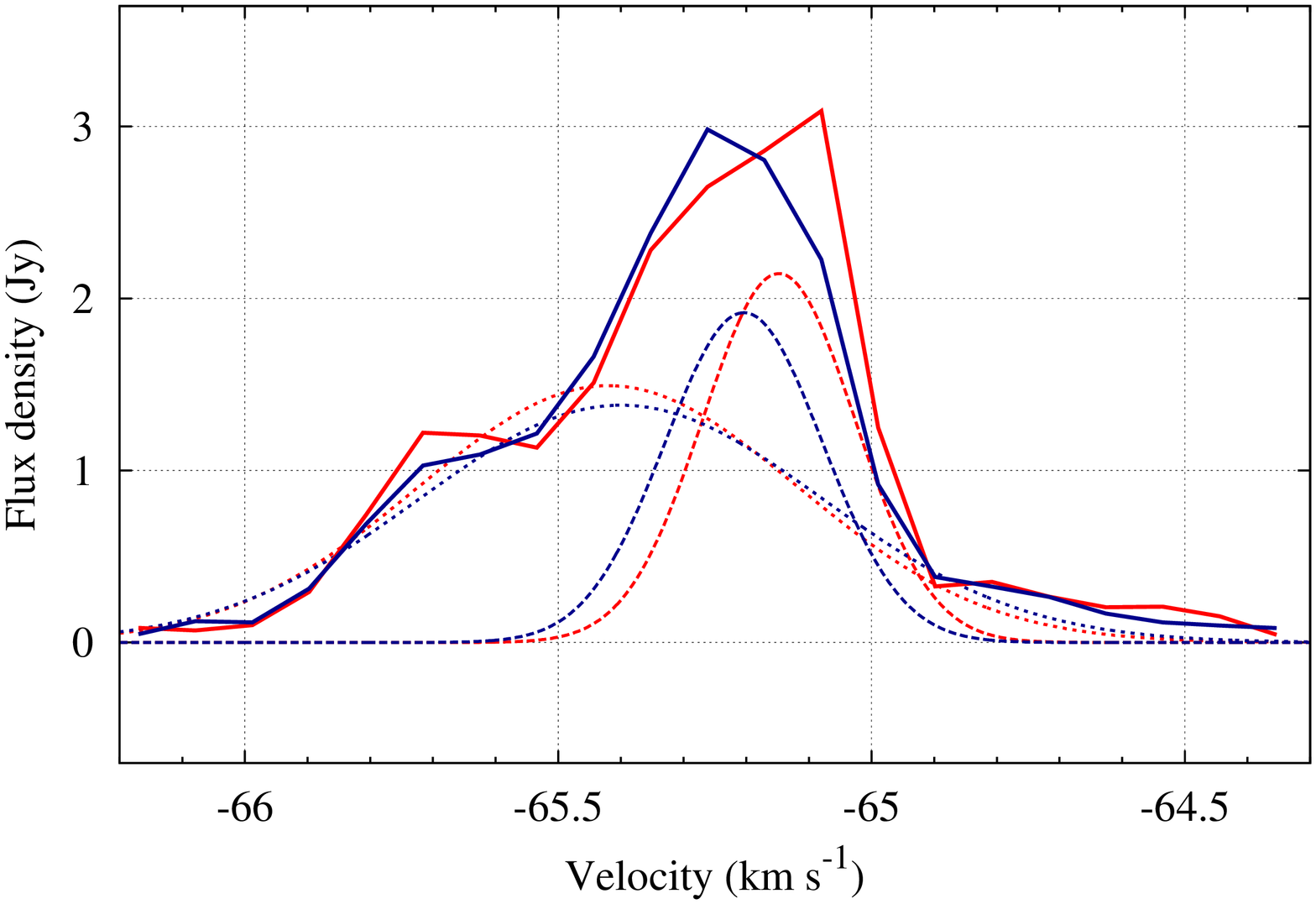}
\caption{MERLIN 1612-MHz spectra of the left (red solid line) and right (blue solid line) circular polarizations 
         of maser features of close spatial coincidence. The dashed lines show the Gaussian curves fitted to the 
         data. The strongest component near $-$65.18\,km\,s$^{-1}$ exhibits a velocity shift between the left 
         (red dashed line) and right (blue dashed line) circular polarizations, which is a signature of Zeeman 
         splitting.\label{zeeman_pair}}
\end{figure}

Two groups of spatially coincident ($<$0\farcs1) RHC and LHC features at 
RA(J2000) = 01$^{\rm h}$33$^{\rm m}$51\fs2247, Dec(J2000) = 62\degr26\arcmin53\farcs311 appear in 8 contiguous channels. 
These are marked by $z$ in Table \ref{circular}. The features are clearly asymmetric but can be easily fitted with two slightly
shifted Gaussian components (Fig.\ref{zeeman_pair}). The strongest component near $-$65.18\,km\,s$^{-1}$ shows 
a 0.142$\pm$0.045\,km\,s$^{-1}$ difference in the peak velocities between the RHC and LHC emissions.
If one can interpret this difference as a Zeeman effect, then the magnetic field strength is $-$0.60$\pm$0.19\,mG
assuming an average splitting coefficient of the $\sigma$ components \citep{davis74}. 
Here, the minus sign indicates that the direction of the field is towards the Earth. 
No linearly polarized emission is detected at this position.

\subsection{Emission from the near and far caps}
The 1612-MHz emission in the velocity ranges from $-$66.8 to $-$66.0\,km\,s$^{-1}$ ($b_{\rm cap}$) and from $-$43.6 to 
$-$42.8\,km\,s$^{-1}$ ($r_{\rm cap}$) is unresolved or barely resolved with the MERLIN beam. The lower limits to the brightness
temperature ($T_{\rm b}$) for $b_{\rm cap}$ and $r_{\rm cap}$ regions are 0.7$-$3.1$\times$10$^{7}$ and 3.3$-$6.3$\times$10$^{7}$\,K, 
respectively. $T_{\rm b}$ of the emission from $b_{\rm cap}$ is a factor of 2$-$7 higher than that of $r_{\rm cap}$, 
while the angular size is by 18$-$23\% lower. Both extreme regions coincide within 70\,mas supporting the standard model
of spherically symmetric shell. 
 The emission from the near cap may contain a component from the amplified stellar image 
(\citealt{norris84}; \citealt*{vanlange93}; \citealt*{vanlange00}). We failed to verify this hypothesis
due to the insufficient angular resolution of our data and the poorly known optical position of the underlying star.

\section{Discussion}
\subsection{Revised properties of OH127}
The present observations allow us to refine the basic stellar parameters of OH127. \citet*{whitelock91} derived 
a period-luminosity, $P -L_{\star}$ relation for OH/IR stars with pulsation periods of 1000$-$2000\,days. Using their 
 eq. 2, one would predict
 that the luminosity of OH127 is $2.68^{+1.81}_{-1.08} \times 10^4 L_{\sun}$. The uncertainty is rather large but this
value is close to $1.5\times 10^4 L_{\sun}$, which is the value adopted by \citet{debeck10} scaled to our distance estimate. 
\citet{kemper02} obtained the best fit of the spectral energy distribution (SED) of OH127 to the observations 
at the distance of 1.8\,kpc for the assumed luminosity $0.63\times 10^4 L_{\sun}$ which translates to $2.9\times 10^4 L_{\sun}$ 
for our distance estimate (Tab. \ref{maintable}).Thus, the stellar luminosity of OH127 is below the AGB-limit of 
$5.5\times 10^4 L_{\sun}$ \citep{wood83} but it is very luminous compared to a typical AGB star.

 OH127 emits virtually all its radiation at infrared wavelengths and the mass-loss rate, $\dot M$, obtained
from modelling of the spectral energy distribution (SED) is 1.5$-$2.8$\times$10$^{-4}$M$_{\sun}$\,yr$^{-1}$  (\citealt{justtanont92};
\citealt{kemper02}) when scaled to our distance estimate and for an assumed dust to gas ratio of 0.01.
The observations and modelling of multiple low- and high$-J$ CO lines (up to J = 7$-$6) by \citet{debeck10} 
give
$\dot M$ = 3.4$\times$10$^{-5}$M$_{\sun}$\,yr$^{-1}$ for the distance of 3.87\,kpc. 
The mass-loss rate calculated from an empirical formula based on the photo-dissociation model 
\citep{netzer87} is 7.1$\times$10$^{-5}$M$_{\sun}$\,yr$^{-1}$.  This value is only a factor of two higher
than that obtained with CO observations \citep{debeck10} and is typical for OH/IR stars.
This indicates that the $\dot M$ estimates from the SED modelling are an order of magnitude higher
than those from CO and OH data,
likely due to the poorly constrained dust to gas ratio \citep{justtanont92}.

\subsection{Magnetic field}
The strength of the magnetic field tentatively detected in OH127 is comparable in magnitude with that estimated for 
other stellar OH maser sources (\citealt{szymczak98}; \citealt{bains03};\,\cite{amiri10}; 
\citealt{etoka10}). However, it is an order of magnitude higher than that measured from 
single dish high resolution spectra of several OH/IR objects \citep{zell91}. This discrepancy is likely 
due to spatial blending of individual maser clumps in the envelopes and is confirmed for OH127:  our NRT 1612-MHz spectra 
in the velocity range of $-$66.6 to $-$64.5\,km\,s$^{-1}$ contain features with m$_{\rm C}\le$5\%  while the MERLIN maps 
show highly (up to 38\%) polarized components (Tab. \ref{circular}).

The presence of magnetic field in the outer envelope of OH127 is implied by the net polarization observed in both
OH transitions. Specifically, the source shows a strong excess of LHC emission well seen in the NRT spectra and MERLIN maps 
(Figs. \ref{spectr1612} $-$ \ref{map1667}). Enhancing one sense of circular polarization and inhibiting the other can be 
explained by the overlap of different Zeeman components due to a velocity gradient within the medium 
\citep{deguchi86} or matching the gradients of the magnetic field and velocity \citep{cook66}.
The latter mechanism may be less probable in a thin circumstellar shell. A simple model of dipole magnetic field with an adapted
Cook's mechanism was successful in explaining the spatial segregation of the 1612-MHz spots dominated by the two senses
of circular polarization in the supergiant star VX\,Sgr \citep{zell96}. It is quite possible that Cook's mechanism
could also partly explain the distribution of the 1612-MHz maser polarized components of OH127 (see Fig. \ref{map1612}) by
adopting a scenario inspired from Zell \& Fix's Fig. 4 but with dipole field lines stretched in the radial direction and 
a different inclination of the dipole axis to the line of sight.
At a distance of 5350\,au from the star a radial magnetic field or a field with open lines can occur. 
In the first case the outflow velocity gradient can combine with the Zeeman pair to produce a coherence length of maser amplification 
longer for LHC component than for RHC component at the near and far sides of the envelope, i.e. the blue- and red-shifted parts 
of the spectrum are dominated by LHC emission. If the field with open lines or of interstellar origin is oriented towards the observer, 
as shown by our observations, then the masers from the near and far sides of the envelope will be dominated by LHC and RHC emission, 
respectively. 
This scenario is not fully consistent with our data where the red wing shows positive and negative values of $m_{\rm C}$ at
1612\,MHz and only negative values at 1667\,MHz.

OH127 at 1612\,MHz shows very small differences in the polarization angles at the near and far sides of the shell, suggesting a regular
configuration of the magnetic field. Such a phenomenon was observed in other OH/IR objects and two explanations are proposed in
\citet{wolak12}: (1) the field is of stellar origin and the same orientation of the field is carried away by the wind
flowing towards the near and back sides of the shell, (2) the field is of interstellar origin and can be amplified and/or distorted 
by the stellar wind. 

In the first case the field of 0.6\,mG strength of solar type ($B\sim r^{-2}$) gives a surface field strength
of $\sim$1kG for the assumed star radius of 859$\mathrm R_{\sun}$ \citep{debeck10}, which is $\le 10^3$ times 
stronger than the average magnetic field ($\sim$few G) for an AGB star \citep{kemball91}. 
 In the second case, if the magnetic field pressure equals the gas kinetic pressure, then  $B^2/8\pi \sim \rho V_{\rm e}^2$, 
where $\rho$ is the gas density in g\,cm$^{-3}$.
 Since $\rho =  \dot M /4\pi r^2 V_{\rm e}$ for a free radial expansion then  
$B$(G) $\sim (2 \dot M V_{\rm e})^{0.5}/r$. With $\dot M$ in units of $10^{-5}M_{\sun}$\,yr$^{-1}$, $V_{\rm e}$ in units of 10\,km\,s$^{-1}$ and
the distance to the star, $r$ in au we obtain 

\begin{equation}
B(\mathrm G) = 2.4\times(2 \dot M V_{\rm e})^{0.5}/r.  
\end{equation}

\noindent
For $\dot M$ = 3.4 \citep{debeck10}, $V_{\rm e}$ = 1.12, $r$=5350\,au (Table \ref{maintable}) this equation
yields a B field strength of 0.9\,mG, which is close to our value of 0.6\,mG. This means that in our measured OH maser 
region there is nearly equilibrium between the magnetic and kinematic pressure. Eq. 1 may not be true close to the star where 
a solar ($\sim r^{-2}$) or dipole ($\sim r^{-3}$) type field may exist. 
On the other hand, the magnetic field measurements from maser observations indicate that the field strength follows 
the relation $B \sim n^{0.5}$ over an enormous range of the number density, $n$, \citep{vlemmings12}. 
This empirical relation predicts a number density of molecular hydrogen of $1\times10^5$\,cm$^{-3}$ for a 0.6\,mG field.   
For the kinematic model \citep{sun87} such a density is 
sufficient to sustain the 1612-MHz maser operation when $\dot M> 5\times 10^{-5}$ and $>1\times 10^{-5}$M$_{\sun}$\,yr$^{-1}$ for
relative abundances of OH molecule higher than $2\times 10^{-4}$ and $5\times 10^{-4}$, respectively. These mass-loss rate 
values are consistent with those recently obtained with the CO observations and modelling (\citealt{debeck10} and Sect. 4.1).

We note that a magnetic field of 0.6\,mG is two orders of magnitude larger than the average interstellar magnetic field 
\citep{troland86}. This suggests that the magnetic field in the OH regions could be compressed or amplified
by a factor of $\sim$100 \citep{soker02}. A model of field compression by \citet{gerard85} predicts an amplification
factor of up to 10. Another possibility is local enhancements of the magnetic field related to 
the stellar spots or convective cells which might produce the magnetic clouds similarly as in the solar wind \citep{soker06}.


\subsection{Saturation level}
Our observations show that the lower limit to the brightness temperature of the strongest 1612-MHz emission is 
up to 6$\times 10^7$\,K (Sect. 3.6). VLBI observations at a phase of 0.67 detected only the extreme blue-shifted 
emission near $-$66.0\,km\,s$^{-1}$, which was unresolved with $\sim$100\,mas resolution, while the extreme red-shifted emission 
was resolved out \citep{norris84}. $T_{\rm b}$ for the unresolved emission was estimated at  
$5\times10^9$\,K. Our MERLIN data were obtained at phase 0.44, so it is likely that the flux density level was 
similar to that observed with the VLBI. Using the VLBI estimates of size and brightness temperature for the extreme 
blue-shifted emission and the distance and shell radius given in Table \ref{maintable} we estimated that the maser beaming 
solid angle, $\Omega$, is 0.016\,sr. This is close to the canonical value \citep{goldreich72}. 
From our comparison of the angular sizes of the two extreme maser regions (Sect. 3.6) we get 0.022$-$0.024\,sr 
for the beaming solid angle of the red-shifted emission.
In the standard maser theory \citep*{goldreich73} the 1612\,MHz maser stimulated emission rate 
can be approximated \citep{amiri10}) by

\begin{equation}
 R = 1.3\times10^{-2}\left (\frac{T_{\rm b}}{10^{10}\mathrm{K}}\right)\left (\frac{\Omega}{10^{-2}\mathrm{sr}}  \right ) {\mathrm s}^{-1}
\end{equation}

\noindent
This implies $R$ = 0.011\,s$^{-1}$ and 0.008$-$0.002\,s$^{-1}$ for the blue- and red-shifted emission regions, respectively.
We notice that the first value is consistent with $R$ predicted for interstellar OH masers \citep{goldreich73}.
The radiative and collisional decay rate for the 1612\,MHz is not observationally constrained but its upper limit
can be estimated \citep{elitzur92} with the following equation

\begin{equation}
 \Gamma < 2.7\times10^{-2}\left (\frac{T_{\rm b}}{10^{10}\mathrm{K}}\right)\left (\frac{\Omega}{10^{-2}\mathrm{sr}}  \right ) {\mathrm s}^{-1}
\end{equation}

\noindent
With the above estimated values of $T_{\rm b}$ and $\Omega$ for the blue-shifted emission $\Gamma < 1.5\times 10^{-3}$\,s$^{-1}$
which implies $R/\Gamma>$7, i.e. the maser in OH127 is weakly saturated. 
 This result appears to be consistent with previous observations showing a fixed relation between 
the OH and infrared emissions \citep{harvey74}. The lack of relatively large random variations in the OH maser fluxes 
of OH127 also suggests that the maser emission is saturated.

\section{Summary and conclusions}
Full polarization monitoring of 1612- and 1667-MHz OH maser lines over a period of 6.5\,years and one epoch polarimetric
map of OH127 led to the following new results: 

(1) The OH maser fluxes show large (0.8$-$1.4\,mag) amplitude variations with period of $\sim$1600\,days. The ratios of 
the integrated flux at 1612 and 1667\,MHz lines of 5.4 and 9.9 at the minimum and maximum, respectively indicate 
a partial suppression of the 1667\,MHz emission.

(2) The comparison of the light-travel diameter of the 1612-MHz maser shell obtained using the NRT with the angular diameter
determined from MERLIN maps yields a distance of 3.87$\pm$0.28\,kpc.

(3) The fluxes of polarized emission follow the variations of the total flux, whereas the degrees of circular and linear 
polarization are constant within measurement accuracy.

(4) The tentative detection of a Zeeman pair at 1612\,MHz implies a magnetic field strength of $-$0.6\,mG at a distance of 
$\sim$5400\,au. The strong alignment of polarization position angles at both the near and far sides of the shell and the presence 
of a net polarization suggest a regular magnetic field.

(5) The stellar luminosity of OH127 is slightly below the AGB limit and the mass loss rate is typical for an AGB star.

(6) The measured sizes of the compact emissions from the near and far sides of the shell and the lower limits to the brightness 
temperature imply weakly saturated emission.   
  
\section*{Acknowledgments}
P.W. acknowledges support by the European Union PhD scholarship programme ZPORR.
The Nancay Radio Observatory is the Unit\'e Scientifique de Nan\c{c}ay of 
the Observatoire de Paris, associated with the CNRS. The Nan\c{c}ay Observatory
acknowledges the financial support of the R\'egion Centre in France.
MERLIN is a UK national facility operated by the University of Manchester on behalf of STFC.




\begin{table*}

\caption{Highly circularly polarized components in OH127.8+0.0. The columns are as follows:
  (1) RA and DEC relative to the shell centre (Table \ref{maintable}); 
(2) the peak Stokes $I$ flux density, together with the rms $(1\sigma)$;
  (3) the peak Stokes $V$ flux density, together with the rms $(1\sigma)$; (4) the degree of circular
  polarization, together with the rms $(1\sigma)$; (5) the peak $p$ flux density, together with the rms $(1\sigma)$;
  (6) the degree of linear polarization, together with the rms $(1\sigma)$; 
  (7) the LSR velocity of the component; (8) Zeeman pair.}\label{circular}
\begin{tabular}{r r r r r r r c c}
\hline
\multicolumn{1}{c}{RA}&\multicolumn{1}{c}{Dec} &\multicolumn{1}{c}{$I(\sigma_I)$} &
\multicolumn{1}{c}{$V(\sigma_V)$} &\multicolumn{1}{c}{$m_{\mathrm C}(\sigma_{m_{\mathrm C}})$}&
\multicolumn{1}{c}{$p(\sigma_p)$} &\multicolumn{1}{c}{$m_{\mathrm L}(\sigma_{m_{\mathrm L}})$}&
\multicolumn{1}{c}{$v_{LSR}$} & \multicolumn{1}{c}{notes}\\ 
\multicolumn{2}{c}{(mas)}&
\multicolumn{2}{c}{(mJy\,b$^{-1}$)}&
\multicolumn{1}{c}{(\%)} &
\multicolumn{1}{c}{(mJy\,b$^{-1}$)} &
\multicolumn{1}{c}{(\%)} &
\multicolumn{1}{c}{(km\,s$^{-1}$)} &  \\
\multicolumn{2}{c}{(1)}&\multicolumn{1}{c}{(2)}& \multicolumn{1}{c}{(3)}&
\multicolumn{1}{c}{(4)}& \multicolumn{1}{c}{(5)}&
\multicolumn{1}{c}{(6)}&
\multicolumn{1}{c}{(7)}&
\multicolumn{1}{c}{(8)} \\ 
\hline
\multicolumn{9}{c}{\it 1612\,MHz} \\
\hline
      $-$200       & $-$240     &  3145(23)  &  $-$213(21)   &  $-$6.8(0.7)&           &          &  $-$66.63 &  \\
      $-$80        & $-$160     &  1016(21)  &  $-$183(18)   &  $-$18.0(2.1)&           &          &  $-$66.53 &  \\
      $-$320       & $-$280     &  3030(19)  &  $-$221(18)   &  $-$7.3(0.6)& 63(12)  &2.1(0.4)  &  $-$66.40 & \\
      $-$200       & $-$120     &  3201(24)  &  $-$168(18)   &  $-$5.2(0.6)& 92(12)  &2.9(0.4)  &  $-$66.17 & \\
      $-$60        & $-$70      &  22260(49) &  $-$1870(27)  &  $-$8.4(0.2)& 617(13)  &2.8(0.1)  &  $-$65.76 & \\
         160       &  320       &  1250(24)  &  $-$160(20)   &  $-$12.8(1.8)&           &       &  $-$65.44 &  \\
         10        &   70       &  2651(20)  &  $-$413(18)   &  $-$15.6(0.6)&           &       &  $-$65.08 & z \\
      $-$640       &  400       &  534(20)   &  $-$111(16)   &  $-$20.8(3.8)&           &       &  $-$64.26 &   \\
         720       &  480       &  413(18)   &  $-$125(16)   &  $-$30.2(5.2)&           &       &  $-$63.72 &   \\
         860       &   20       &  244(18)   &  $-$93(16)    &  $-$38.1(9.4)&           &       &  $-$62.95 &   \\
         920       &  200       &  482(18)   &  $-$112(16)   &  $-$23.2(4.2)&           &       &  $-$62.36 &   \\
         760       &   40       &  279(19)   &  $-$104(16)   &  $-$37.2(8.3)&           &       &  $-$62.27 &   \\
                   &            &           &                &               &           &       &           &  \\
     $-$200        &  $-$480    &  1463(24)  & 121(18)       &   8.3(1.4)&           &       &  $-$66.17 &  \\
     $-$500        &  $-$360    &  1345(25)  & 126(18)       &   9.4(1.5)&           &       &  $-$66.13 &  \\
     $-$100        &   80       &  2808(21)  & 139(18)       &   4.9(0.7)&           &       &  $-$65.26 & z \\

      440          &  640       &  724(20)   & 188(16)       &   26.0(3.0)&           &        &  $-$64.54 &  \\
                   &            &              &             &               &           &        &           &  \\
     $-$200        &  880       &  369(20)   & $-$131(16)    &  $-$35.5(6.3)&           &        &  $-$47.28 &  \\
     620           &  620       &  420(19)   & $-$119(18)    &  $-$28.3(5.6)&           &        &  $-$46.96 &  \\
     $-$250        &  430       &  1179(20)  & $-$116(16)    &  $-$9.8(1.5)&           &        &  $-$44.92 &  \\
     $-$60         & $-$760     &  912(20)   & $-$122(18)    &  $-$13.4(2.3)&           &        &  $-$44.70 &  \\
     $-$240        &  300       &  1073(20)  & $-$130(17)    &  $-$12.1(1.8)&           &        &  $-$44.24 &  \\
     $-$480        &  0.0       &  1754(20)  & $-$115(18)    &  $-$6.5(1.1)&           &        &  $-$43.83 &  \\
     $-$360        &  $-$500    &  1158(18)  & $-$116(17)    &  $-$10.0(1.6)&           &        &  $-$43.60 &  \\
     $-$80         &  $-$120    &  1721(19)  & $-$102(18)    &  $-$5.9(1.1)&           &        &  $-$43.47 &  \\
                   &            &            &               &               &           &        &           &  \\
     380           & $-$750     &  520(19)   & 141(17)       &  27.1(4.2)&           &        &  $-$46.15 &  \\
     310           & $-$720     &  663(19)   & 183(17)       &  27.6(3.3)&           &        &  $-$45.69 &  \\
     320           & $-$640     &  600(19)   & 124(16)       &  20.7(3.3)&           &        &  $-$45.28 &  \\
     520           & $-$280     &  942(20)   & 125(16)       &  13.3(2.0)&           &        &  $-$44.92 &  \\
                   &            &            &               &               &           &        &           &  \\
\hline
\multicolumn{9}{c}{\it 1667\,MHz} \\
\hline
     $-$240        &  200       &  715(19)    & $-$173(23)   &  $-$24.2(3.8)&          &            &  $-$44.00 &  \\
     $-$480        &    0       &  711(21)    & $-$178(24)   &  $-$25.0(4.1)& 93(16)  &13.0(2.6)  &  $-$43.98 & \\
        0          &  $-$40     &  1490(20)   & $-$209(22)   &  $-$14.0(1.7)& 90(16)  &6.0(1.5)  &  $-$43.85 &  \\
     $-$60         &  $-$80     &  2695(24)   & $-$326(23)   &  $-$12.1(1.0)& 90(16)  &3.3(0.6)  &  $-$43.72 &  \\
     $-$120        &  $-$280    &  1684(23)   & $-$247(22)   &  $-$14.7(1.5)&          &            &  $-$43.58 & \\
\hline
\end{tabular}

\end{table*}


\begin{table*}
\caption{Highly linearly polarized components in OH127.8+0.0. The columns are as follows:
  (1) RA and DEC relative to the shell centre (Table \ref{maintable}); 
 (2) the peak Stokes $I$ flux density together with the rms $(1\sigma)$; 
  (3) the $p$ peak flux density together with the rms $(1\sigma)$; 
  (4) the degree of linear polarization together with the rms $(1\sigma)$; 
  (5) the polarization position angle together with the rms $(1\sigma)$;
  (6) the peak Stokes $V$ flux density together with the rms $(1\sigma)$; 
  (7) the degree of circular polarization together with the rms $(1\sigma)$; 
  (8) the LSR velocity of the component. }\label{linear}
\begin{tabular}{r r r r r r r r c}
\hline
\multicolumn{1}{c}{RA}&\multicolumn{1}{c}{Dec} &\multicolumn{1}{c}{$I(\sigma_I)$}&
\multicolumn{1}{c}{$p(\sigma_p)$}& \multicolumn{1}{c}{$m_{\mathrm L}(\sigma_{m_{\mathrm L}})$}&
\multicolumn{1}{c}{$\chi(\sigma_{\chi})$} & 
\multicolumn{1}{c}{$V(\sigma_V)$} &\multicolumn{1}{c}{$m_{\mathrm C}(\sigma_{m_{\mathrm C}})$}& 
\multicolumn{1}{c}{$v_{LSR}$}  \\ 
\multicolumn{2}{c}{(mas)} &
\multicolumn{2}{c}{(mJy\,b$^{-1}$)} &
\multicolumn{1}{c}{(\%)} &\multicolumn{1}{c}{(\degr)}&
\multicolumn{1}{c}{(mJy\,b$^{-1}$)}  &
\multicolumn{1}{c}{(\%)} &
\multicolumn{1}{c}{(km\,s$^{-1}$)}   \\
\multicolumn{2}{c}{(1)}&\multicolumn{1}{c}{(2)}& \multicolumn{1}{c}{(3)}& \multicolumn{1}{c}{(4)}& 
\multicolumn{1}{c}{(5)}& \multicolumn{1}{c}{(6)}& \multicolumn{1}{c}{(7)} & \multicolumn{1}{c}{(8)}\\

\hline
\multicolumn{9}{c}{\it 1612\,MHz} \\
\hline
   $-$180 &$-$340  & 2733(21)  & 108(12)  &  4.0(0.5)  &  4.7(1.9)&               &               & $-$66.63  \\
   $-$280 &$-$140  & 3701(21)  & 136(12)  &  3.7(0.3)  &  40.0(1.7)&$-$128(18) & $-$1.7(0.25)  & $-$66.35  \\
   $-$360 &$-$20   & 2354(26)  & 132(12)  &  5.6(0.6)  &  $-$0.2(5.1)&               &               & $-$66.08  \\
   $-$45  &$-$75   & 19747(42)  & 505(13)  &  2.5(0.9)  &  $-$15.5(1.0)&$-$11013(27) & $-$7.0(0.1)  & $-$65.76  \\
   $-$40  &$-$40   & 1461(26)  & 137(11)  &  9.4(0.9)  &  0.6(1.7)&               &               & $-$65.35   \\
   $-$120 &  200   & 4396(19)  &  86(11)  &  2.0(0.3)  &  $-$37.6(2.2)& 252(18) & 5.7(0.4)  & $-$65.26  \\
          &        &             &           &              &                &               &               &             \\
\hline
\multicolumn{9}{c}{\it 1667\,MHz} \\
\hline
$-$120    &160     & 773(22)  & 203(16)  &  26.3(2.8)  &  $-$2.4(1.4)&               &               & $-$44.11    \\
$-$520    &$-$80   & 545(20)  & 177(16)  &  32.5(4.1)  &  0.1(1.5)&$-$100(23) &$-$18.3(4.9)   & $-$44.02  \\
$-$460    &$-$280  & 449(23)  & 163(15)  &  36.3(5.2)  &  $-$7.5(1.5)&               &               & $-$43.89   \\
$-$160    &$-$80   & 1127(25)  & 185(16)  &  16.4(1.8)  &  $-$6.1(1.5)&96(23) &8.5(2.2)   & $-$43.80  \\
$-$120    &$-$200  & 2641(26)  & 153(16)  &  5.8(0.7)  &  $-$41.7(1.6)&$-$189(22) &$-$7.1(0.9)   & $-$43.67  \\
$-$120    &$-$80   & 1810(23)  & 125(16)  &  6.9(1.0)  &  $-$36.2(1.8)&$-$165(22) &$-$9.1(1.3)   & $-$43.58   \\
   120    &  0     & 268(18)  & 110(15)  &  41.0(8.3)  &  38.3(2.8)&               &               & $-$43.50    \\
\hline
\end{tabular}
\end{table*}

\label{lastpage}

\end{document}